\documentclass{PoS}

\title{Lattice computation of the quark propagator in Landau gauge at finite temperature}

\ShortTitle{Lattice computation of the quark propagator in Landau gauge at finite temperature}

\author{\speaker{Paulo J. Silva}, Orlando Oliveira\\
        CFisUC, Department of Physics, University of Coimbra, 3004-516 Coimbra,
Portugal\\
        E-mail: \email{psilva@uc.pt}, \email{orlando@uc.pt} }

\abstract{We report on the lattice computation of the quark propagator at finite temperature in the Landau gauge, using quenched gauge configurations. The propagator form factors are computed for various temperatures, above and below the gluon deconfinement temperature $T_c$, and for all the Matsubara frequencies. Our results suggest a strong connection between quark and gluon deconfinement and  chiral symmetry restoration above $T_c$. }

\FullConference{37th International Symposium on Lattice Field Theory - Lattice2019\\
		16-22 June 2019\\
		Wuhan, China}

\begin{document}

\section{Introduction and Motivation}

In recent years, an intense research activity has been performed towards a full understanding of the QCD phase diagram. Following the current experimental programs in several facilities around the world, there has been a plethora of theoretical studies concerning QCD at finite temperature and density.

Quarks and gluons feel their environment and, therefore, their properties depend on the temperature and density of matter of the surrounding bath. At low temperatures, quarks and gluons are confined within hadrons. On the other hand, quarks and gluons behave as free quasiparticles at sufficiently high temperatures or densities.

The QCD two-point correlation functions can provide crucial information about the properties of the fundamental quanta of the theory and,
in particular, about confinement and chiral symmetry breaking.
Following our previous works about the Landau gauge gluon propagator at finite temperature \cite{gluonmass2014, gluonsect2016}, here we focus on the Landau gauge quark propagator. First we resume our recent work \cite{quarkT} and then we also show preliminary results concerning the $Z_3$ dependence of the quark propagator.

\section{Landau gauge quark propagator at finite temperature}

At finite temperature, the  Landau gauge quark propagator in momentum space is described by three form factors  $\omega$, $Z$ and $\sigma$
\begin{eqnarray}
   S(p_4 , \vec{p} ) 
  &=& \frac{ - i \gamma_4 \, p_4 ~ \omega (p_4, \vec{p}) - i \vec{\gamma} \cdot \vec{p} ~ Z(p_4 , \vec{p} ) + \sigma (p_4 , \vec{p} ) }
                                        {   p^2_4 ~  \omega^2 (p_4, \vec{p})   + \left( \vec{p}\cdot\vec{p} \right) ~ Z^2(p_4 , \vec{p} )  + \sigma^2 (p_4 , \vec{p} ) } \label{quarkS}
\end{eqnarray}
that, assuming we are close to the continuum, can be accessed by computing suitable traces of the propagator times $\gamma-$matrices. 

In this work the quark propagator is computed using $\mathcal{O}(a)$ non-perturbative improved Wilson fermions (see \cite{quarkT} for details)
and for the lattice setup described in Table \ref{lattsetup}.  Our calculation focus in the region around the critical temperature $T_c\sim 270$ MeV and uses
quenched configuration ensembles generated with the  Wilson gauge action, with a  spatial physical volume around $(6.5 \textrm{fm})^3$; 
details about the lattice ensembles can be found in \cite{gluonmass2014, quarkT}. For each of the simulations reported here the quark propagator
was computed for 100 configurations and for two point sources per gauge configuration. The exception being the highest temperature considered, were we took only
a single point source per configuration. In all cases, the bare quark mass is around 50 MeV and, in some cases, a quark bare mass around 10 MeV
was also computed. In all cases we average over equivalent momenta. Furthermore, we have applied the cylindrical cut for momenta above 1 GeV, 
ignoring the data whose relative statistical error is above $50\%$.

\begin{table}
  \begin{center}
\begin{tabular}{ccclllcc}
\hline
T     &  $\beta$  &  $L_s^3 \times L_t$  & $\kappa$ & $\kappa_c$ & $a$  & $m_{bare}$ & $c_{sw}$  \\
(MeV) &           &            &         &            & (fm) & (MeV) & \\
\hline
243   & 6.0000     &  $64^3\times8$   & 0.1350 &  0.13520 &0.1016  & 10   & 1.769\\
      &            &                  & 0.1342 &     &  & 53   &      \\
\hline
260   & 6.0347     &  $68^3\times8$   & 0.1351 & 0.13530 & 0.09502  & 11   & 1.734\\
      &            &                  & 0.1344 &  & & 51   &      \\
\hline
275   & 6.0684     &  $72^3\times8$   & 0.1352 & 0.13540 & 0.08974  & 12   & 1.704\\
      &            &                  & 0.1345 & &  & 54   &      \\
\hline
290  & 6.1009     & $76^3\times8$    & 0.1347  & 0.13550  & 0.08502   &  51  & 1.678 \\
\hline
305  & 6.1326     & $80^3\times8$    &  0.1354   &  0.13559 &  0.08077    & 13  & 1.655 \\      
        &                 &                             & 0.1348  &                &            &  53  &   \\
\hline
324   & 6.0000     & $64^3\times6$   & 0.1342  & 0.13520 & 0.1016 &  53   &   1.769 \\      
\hline      
\end{tabular}
  \end{center}\label{lattsetup}
  \caption{Lattice setup.}
\end{table}

A continuum-like  quark wave function
\begin{equation}
  Z_c(p_4, \vec{p})  =  \frac{Z(p_4, \vec{p})}{\omega (p_4, \vec{p})}
\end{equation}
and a running quark mass
  \begin{equation}
   M (p_4, \vec{p})  =  \frac{\sigma(p_4, \vec{p}) }{ \omega(p_4, \vec{p}) } 
  \end{equation}
can be defined by suitable ratios of the form factors defined in Eq. (\ref{quarkS}). 
Note that we use $\omega$ and not $Z$ to define the ratios $Z_c$ and $M$. The choice being due to the impossibility of extracting
$Z(p_4, \vec{p} = 0)$ from the lattice data.

\begin{figure}
\vspace*{-0.3cm}
\begin{center}
\includegraphics[width=0.82\textwidth]{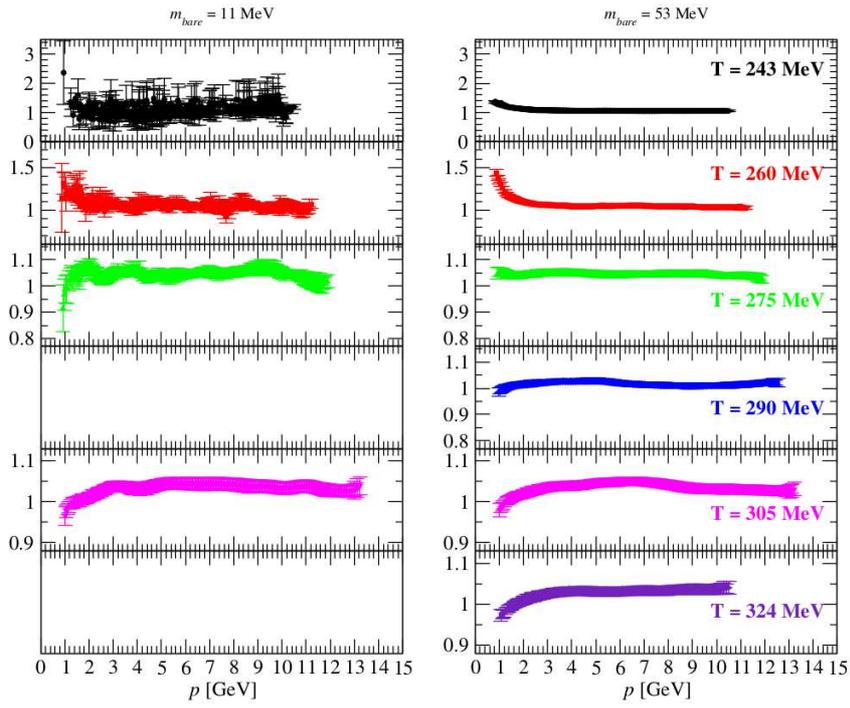}
\end{center}
\caption{Bare quark wave function.}
\label{Zquarkfig}
\end{figure}

\begin{figure}
\begin{center}
\includegraphics[width=0.82\textwidth]{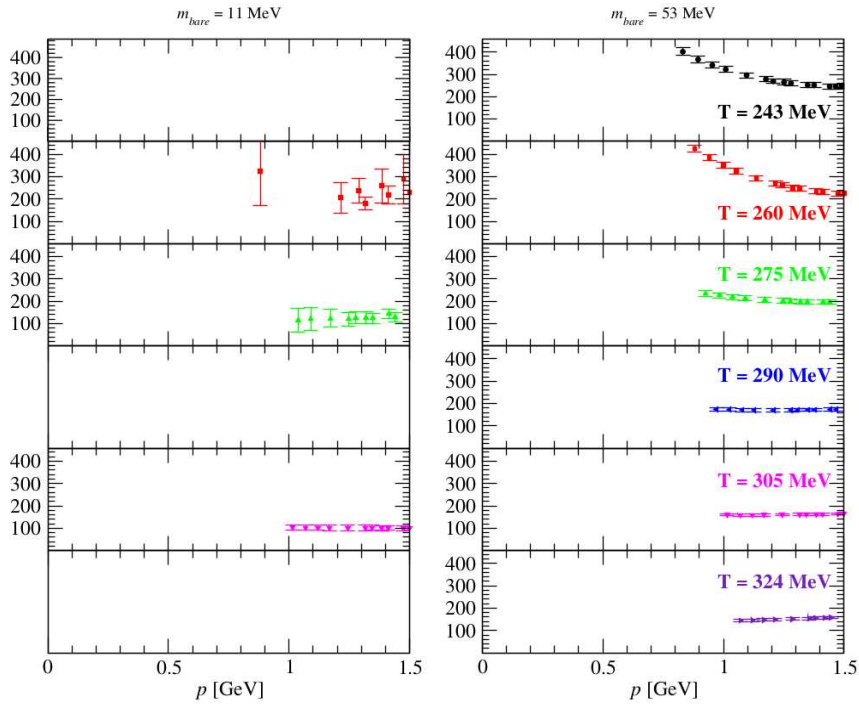}
\end{center}
\caption{Running quark mass.}
\label{Massfig}
\end{figure}

Our results for the quark wave function and for the running quark mass (first Matsubara frequency only) can be seen in Figures \ref{Zquarkfig} and \ref{Massfig}. Note that the lattice data in these figures are not corrected for lattice artefacts. Nevertheless, we see that both quantities are sensitive to the deconfinement phase transition. In particular, the quark wave function (Fig. \ref{Zquarkfig}), at high momenta, approaches a constant from above for $T<T_c$ and from below for $T>T_c$. In what concerns the running quark mass (Fig. \ref{Massfig}) , it decreases with $p$ for temperatures below   $T_c$ and it is essentially flat above the critical temperature.   

\begin{figure}
\vspace*{-0.3cm}
\begin{center}
\includegraphics[width=0.83\textwidth]{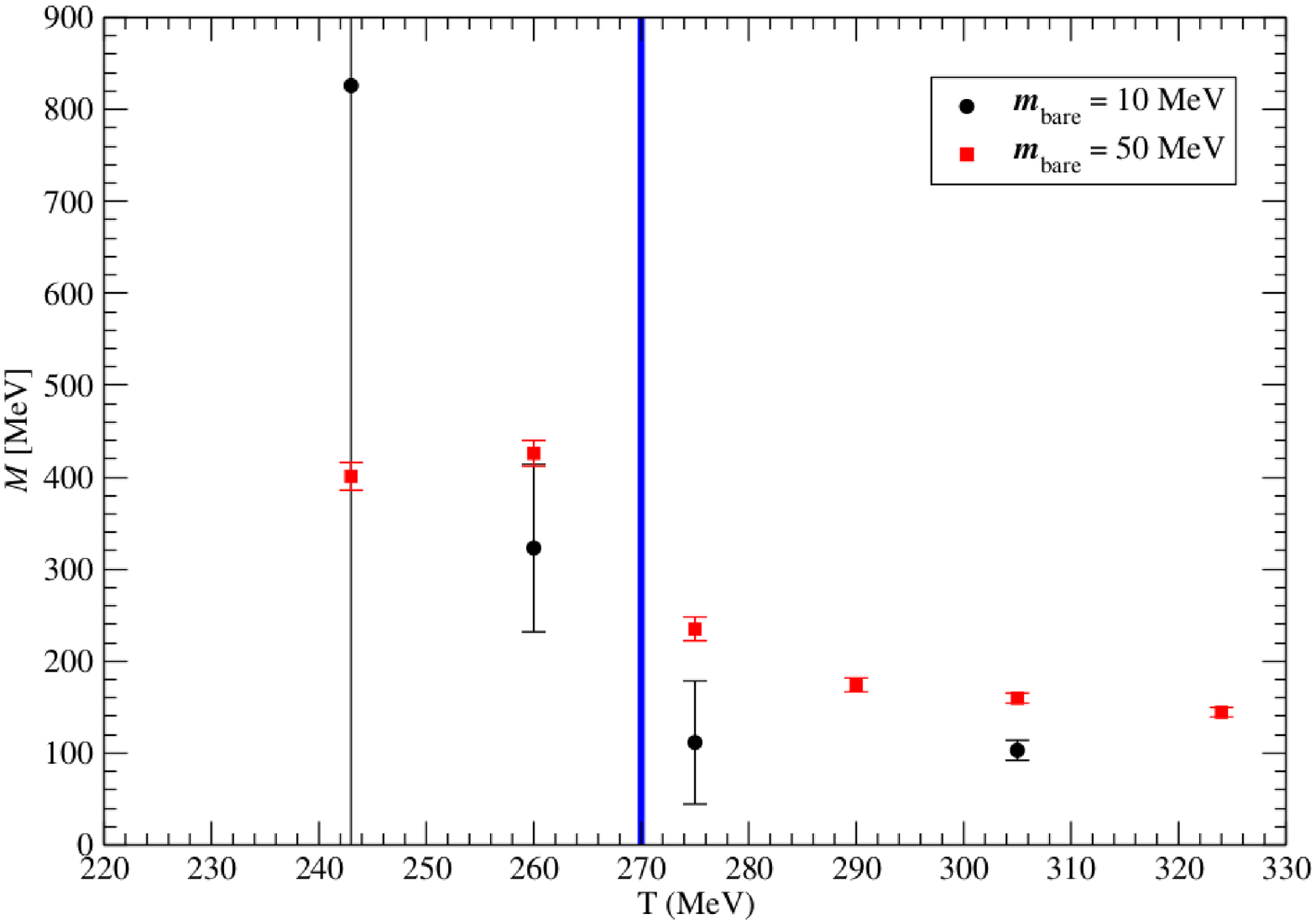}
\end{center}
\caption{ $M(p_4, \vec{p} = 0)$ as function of $T$.}
\label{MassT}
\end{figure}

In Fig. \ref{MassT}, we resume the behaviour of the infrared running mass as a function of the temperature. The lattice data indicates a mass suppression for $T>T_c$, a region where the data favours a quasiparticle scenario, with a mass around $100$MeV.

Further details about this work can be seen in \cite{quarkT}.

\section{$Z_3$ sectors}

In \cite{gluonsect2016}, we have studied how the Landau gauge gluon propagator at finite temperature looks like in different $Z_3$ sectors, i.e.
the phases of the Polyakov loop. 
As reported there, below the critical temperature, the gluon propagator computed in different $Z_3$ sectors are undistinguishable within statistical
errors. However, for $T>T_c$ the gluon propagator for the zero sector decouples from the other sectors. 
The difference between the propagators in different sectors can be used as a criterion to identify the deconfinement phase transition --- see
\cite{gluonsect2016} for further details. 

\begin{figure}
\begin{center}
\includegraphics[width=0.73\textwidth]{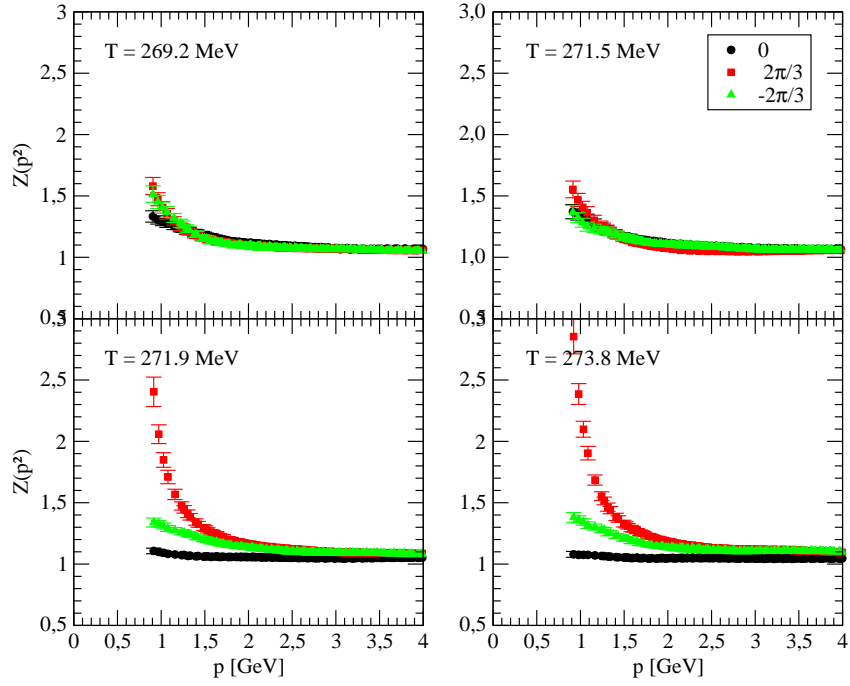}
\end{center}
\caption{Quark wave function $Z(p^2)$ for different $Z_3$ sectors at several temperatures.}
\label{centerZup2}
\end{figure}

\begin{figure}
\vspace*{-0.3cm}
\begin{center}
\includegraphics[width=0.73\textwidth]{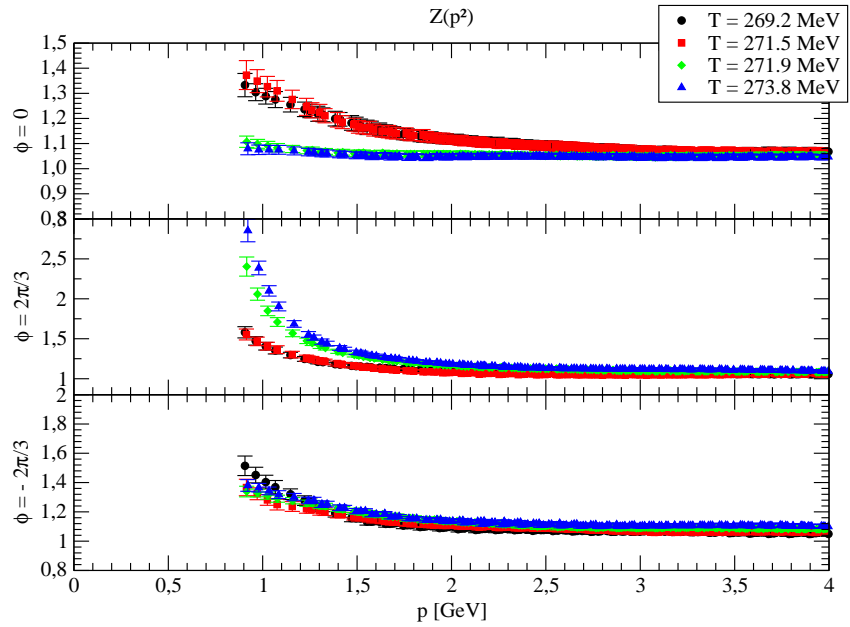}
\end{center}
\caption{Temperature dependence of $Z(p^2)$ for different $Z_3$ sectors.}
\label{centerZsect}
\end{figure}

\begin{figure}
\vspace*{-0.3cm}
\begin{center}
\includegraphics[width=0.73\textwidth]{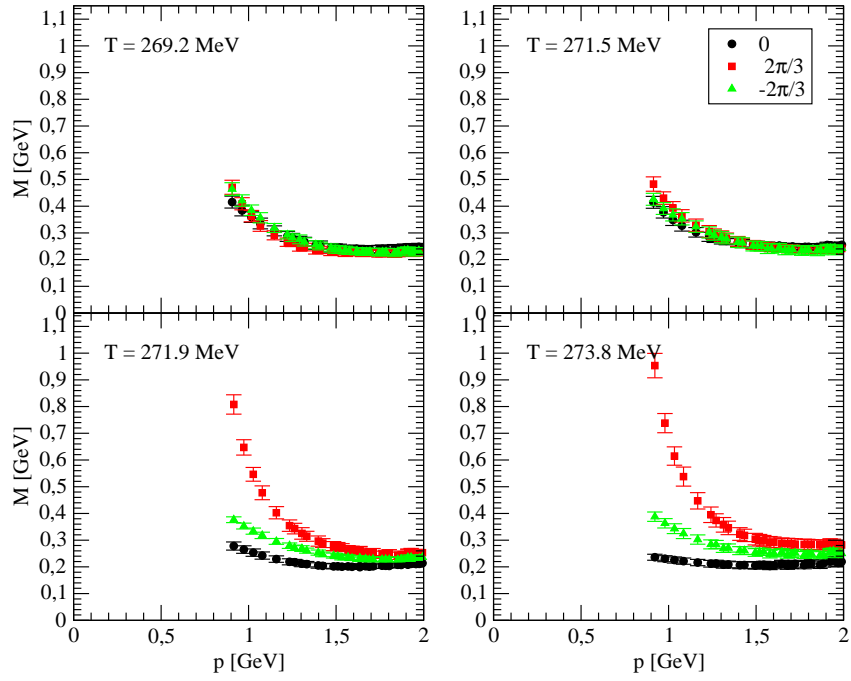}
\end{center}
\caption{Running quark mass for different $Z_3$ sectors at several temperatures.}
\label{centerMup2}
\end{figure}

\begin{figure}
\vspace*{0.2cm}
\begin{center}
\includegraphics[width=0.73\textwidth]{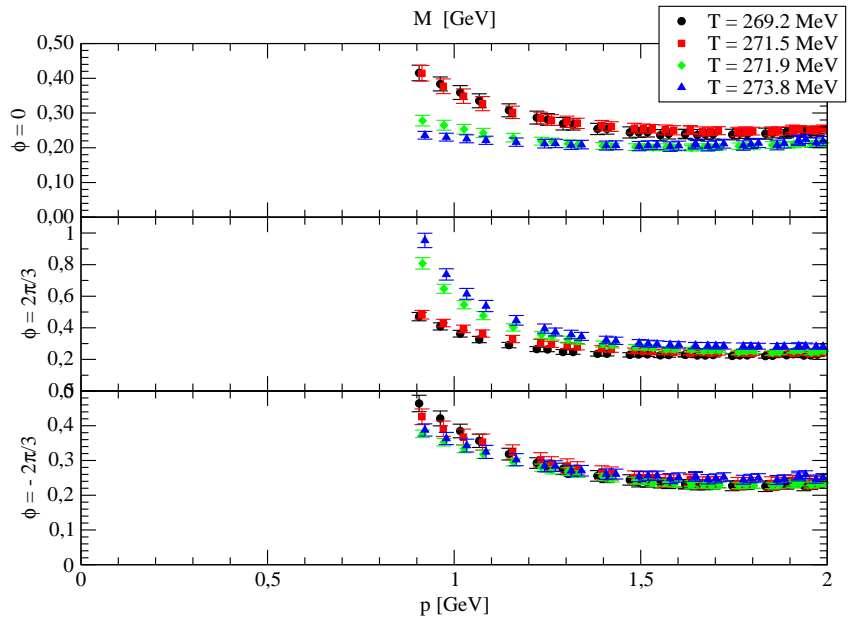}
\end{center}
\caption{Temperature dependence of the running quark mass for different $Z_3$ sectors.}
\label{centerMsect}
\end{figure}

In Figs. \ref{centerZup2} to \ref{centerMsect} we report how the quark propagator form factors depend on the Polyakov loop phase.
In the current investigation only a subset of the ensembles used in  \cite{gluonsect2016} was considered.

For temperatures below $T_c$, no difference between the various $Z_3$ sectors is observed. However, for temperatures above $T_c$,
the various $Z_3$ sectors have quite different propagators. If for the gluon propagator the results associated with Polyakov loop phases 
$\phi=\pm2\pi/3$ were indistinguishable \cite{gluonsect2016}, for the quark propagator the results for $\phi=\pm2\pi/3$ are also different from each other.
Note that the scenario is essentially the same for both form factors.

\section*{Acknowledgements}

The authors acknowledge financial support from FCT under contracts  UID/FIS/04564/2016 and CERN/FIS-COM/0029/2017. P. J. S.  acknowledges partial support by FCT under contracts
SFRH/BPD/40998/2007 and SFRH/BPD/109971/2015. The authors
also acknowledge the Laboratory for Advanced Computing at University of Coimbra (http://www.uc.pt/lca) for providing access to the
HPC resource Navigator. The SU(3) lattice simulations were done using
Chroma \cite{Edwards2005} and PFFT \cite{Pippig2013} libraries.

\end{document}